\documentclass[letter,oldversion]{aa}
\usepackage{txfonts}
\usepackage{graphicx}
\usepackage{color}
\usepackage{amssymb}
\usepackage{epsfig}

\def\aap{A\&A}

\begin{document}

\sloppypar

\title{The appearance of magnetospheric instability 
in flaring activity at the onset of X-ray outbursts in A0535+26}

\author{K.~Postnov\inst{1,2}, R.~Staubert\inst{1}, A.~Santangelo\inst{1}, D.~Klochkov \inst{1}, P. Kretschmar\inst{3},  I.~Caballero\inst{1}}

   \offprints{K. Postnov}

   \institute{
Institut f\"ur Astronomie und Astrophysik, 
University of T\"ubingen, Sand 1, 72076 T\"ubingen, Germany
\and
Sternberg Astronomical Institute, 119999, Moscow, Russia
\and
European Space Agency, European Space Astronomy Center, P.O. Box 78, 28691 Villanueva de la Ca\~nada, Madrid, Spain 
          }
  \date{}
\authorrunning{Postnov et. al.}
\titlerunning{X-ray pre-otburst flares in A0535+26}

   \abstract{
We argue that X-ray flaring variability observed in the transient X-ray pulsar A0535+26 is due to low-mode magnetospheric instability. This instability develops at the onset of accretion, in the thin boundary layer between the accretion disk and neutron star magnetosphere.
As a result, the matter collected in the boundary layer 
can rapidly fall onto the NS surface close to the magnetic poles, 
but not exactly along the field lines by which 
the stationary accretion proceeds. This explains the shift in cyclotron line energy 
measured using RXTE data in a pre-outburst spike, with respect to the line energy 
observed during the main outburst. 
Furthermore, the instability can account for the difference in pulse profiles, and their energy evolution that is different in the pre-outburst flare and main outburst.

   }
\maketitle

\section{Introduction}

The transient X-ray binary pulsar A0535+26 was discovered by the 
\textit{Ariel V} satellite during a giant 
outburst (\cite{coe:1975, rosenberg:1975}). The neutron star (NS) rotates with a
period $P\simeq 103$ s and is in an eccentric ($e=0.47$) orbit about 
a massive O9.7IIIe star  with 
an orbital period of $P_{orb}\sim 110$ d (\cite{coe:2006}). 
The X-ray activity of this transient 
source is complicated: several giant outbursts were observed in 1975, 1980, 1983, 1989, 1994 and 2005, with weaker outbursts sometimes observed at successive periastron passages. 
This type of X-ray activity is common for Be/X-ray binary systems, which represent the largest subclass of massive X-ray binaries. The giant (type II, according to \cite{stella:1986}) outbursts can take place
at any orbital phase and most probably are triggered by the enhanced activity of the optical star (\cite{coe:2006}).
More regular, weaker type I outbursts are associated with enhanced interaction of the neutron star with a circumstellar disk surrounding  the Be-companion (\cite{okazaki:2001}).
A0535+26 did not show outbursting activity between 1994 and 2005.
The last giant outburst of A0535+26 occurred 
after a prolonged quiescence period in May-June 2005 (\cite{tueller:2005}). In September 2005, the subsequent normal type I outburst was extensively
studied by RXTE and INTEGRAL (\cite{caballero:2007}, 2008). These observations revealed somewhat unexpected 
features, which we summarize below. 

\textit{Feature 1}. A short (fraction of a day) pre-outburst X-ray spike,
with a luminosity comparable to the outburst maximum, 
is observed during the rise of the outburst. 

\textit{Feature 2}. The NS pulse period at the onset 
of the outburst, remained constant within the margins of error,
but rapidly 
decreased after the periastron passage.

\textit{Feature 3}. RXTE data shows 
that the energy of the cyclotron resonance scattering feature (CRSF) during the spike
at 90\% confidence is $52.0^{+1.6}_{-1.4}$ keV (\cite{caballero:2008}), which is
notably higher than the value $46.1\pm 0.5$~keV measured 
during the main outburst at the same level of X-ray luminosity.
The latter value is consistent with INTEGRAL measurements  
(\cite{caballero:2007}). 

\textit{Feature 4}. The CRSF energy in the main outburst (within the margins of error)
is independent of the X-ray luminosity. 

\textit{Feature 5}. The X-ray pulse profiles during the spike are different 
from the pulse profiles observed during the main outburst. Pulse profiles during the outburst show an abrupt change at energies above the cyclotron resonance, while those in the pre-outburst spike do not.

The additional inspection of the \textit{Swift} BAT (15-50 keV) light curve of the source 
(Fig. 1) reveals that 
the pre-outburst spike observed by RXTE is only 
one of a collection of flares, with the characteristic time up to a few times $10^4$ s, appearing
at the rising part of the outburst.
The flux evolution then proceeds more smoothly  during the spin-up of the NS
(\textit{Feature 6}), which begins close to periastron passage.
A similar flaring activity was observed during the onset of the preceding giant outburst, at the same flux level, in April 2005, and at the next periastron passage in December 2005.  

\begin{figure*}
 \resizebox{\hsize}{!}{\includegraphics{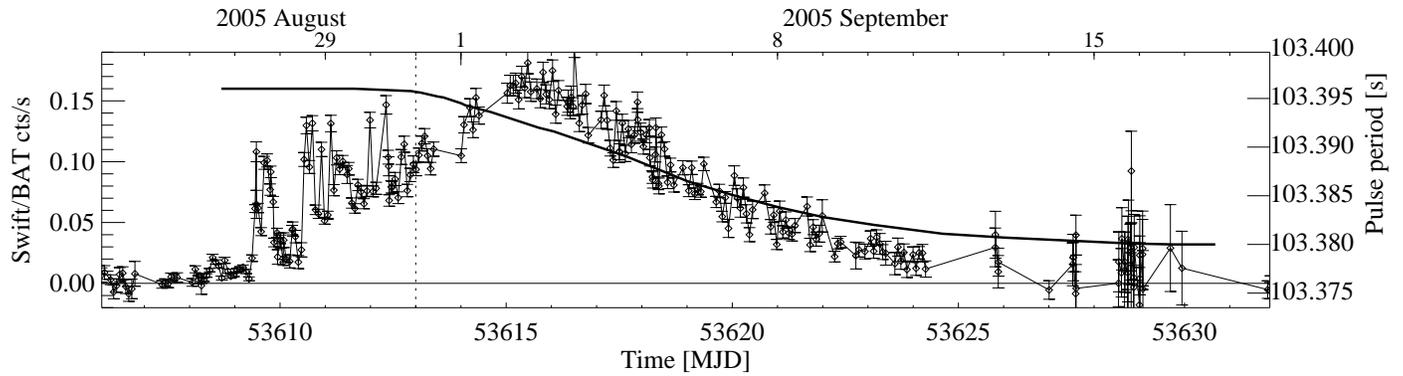}}
 \caption{The Swift/BAT light curve of A0535+26 during the normal
outburst in August/September 2005. The smooth solid line shows the pulse
period development as measured by RXTE (from Caballero et al. 2007b). The dashed line shows the time of periastron passage.}
 \label{batlc}
\end{figure*}

In this Letter we argue that the flaring activity and pre-outburst spikes are most likely 
due to magnetospheric instability. 
This would take place at the onset of a smooth increase in the mass-accretion rate through the accretion disk surrounding the NS, close to periastron passage.
This instability would cause plasma that had accumulated in the thin boundary layer, between the accretion disk and NS magnetosphere, to fall onto the NS surface as large blobs.
These blobs
would be channeled to regions close to the NS magnetic poles by different magnetic field lines than those guiding the accretion flow smoothly increasing from the disk.

\section{The model}

Since the 1970s, it has been known that matter can penetrate the NS magnetosphere, and be accreted onto magnetized NS, via various instabilities.
The gravitational interchange instability (Rayleigh-Taylor,
or Kruskal-Schwarzschild [KS] for plasma in magnetic field) 
of cold plasma at the magnetospheric boundary
(see \cite{elsner&lamb:1976, arons&lea:1976a}), in particular, mediates 
quasi-spherical accretion. 
In this case, close to the magnetic equator where KS instability is more likely, plasma enters the magnetosphere as bubbles or filaments that fall deeper towards the NS.
All or part of these filaments, broken up into smaller pieces 
by Kelvin-Helmholtz instabilities (\cite{arons&lea:1976a}),  
can then be entrained 
by the magnetic field and channeled 
toward the magnetic polar region on the NS surface. 

During disk accretion onto a rotating magnetized 
NS the Kelvin-Helmholtz instability between the disk material and the magnetosphere is also important (\cite{anzer&boerner:1980}, 1983). 
The transition between the accretion disk and the rotating NS magnetosphere (for the aligned magnetic dipoles) 
was studied by Ghosh \& Lamb (1978, 1979), Scharlemann (1978), Anzer \& B\"orner (1983) and
later by Lovelace et al. (1995) using more realistic assumptions.  In the latter model,
the twisting of the stellar magnetic field by the differentially rotating disk leads
to the appearance of open magnetic field lines extending outward from both the star and the disk.
The accretion process may then be maintained by an MHD outflow from the disk that transports away angular momentum.
Numerical MHD simulations of the disk interaction with an inclined magnetic dipole
(\cite{romanova:2003}) confirmed the basic features of 
the disk-magnetospheric interaction. 
Accretion onto NSs can be unstable, leading to outbursts and flares, because of instabilities in the magnetosphere (\cite{baan:1979, spruit:1993}). The KS instability affects accretion onto magnetized NSs at both large and small accretion rates. At small accretion rates, matter accumulates at the boundary layer and flaring events are expected. At large accretion rates, matter is accreted more steadily.

In the case of A0535+26 we  deal with non-stationary accretion 
triggered by enhanced NS interaction with the circumstellar disk and the wind of the Be-companion at binary periastron passage. 
BATSE observations of this source (\cite{bildsten:1997}) 
indicated the strong
disk accretion spin-up during type II (giant) outbursts, while no significant
spin-up was observed during a series of type I outbursts, at successive 
periastron passages both before and after the giant outburst. 
Moreover, the average spin-down of the NS rotation 
is clearly seen in the BATSE data before and
after the giant outburst, suggesting that 
a propeller-like mechanism operates between accretion episodes close to periastron.
We expect similar long-term spin-up/spin-down behavior of magnetized NSs in eccentric Be X-ray binaries on very general grouds (\cite{stella:1986}).
The magnetic propeller regime was investigated by Lovelace et al. (1999). Numerical MHD simulations
(\cite{ustyugova:2006}) revealed the non-stationary character of mass outflow during the propeller
stage. Clearly, the transition from propeller to accretion stage is the most 
complicated for the analysis. Here we suggest that the non-stationary features observed
in the outburst of A0535+26 relate to the magnetospheric instability which develops
during the transition to the accretion stage. 

The pre-outburst X-ray flaring activity is not an exceptional feature of
the 2005 outbursts in A0535+26. Hints of other flares in A0535+26 are apparent for several other type I outbursts in BATSE light-curve data for its giant outburst in 1994 (see Fig. 1 of Finger et al. (1996)).
Similar pre-outburst spikes are observed in 
other Be/X-ray binaries, e.g. in BATSE observations of 
GS 1843-02 (\cite{finger:1999}) and INTEGRAL observations of EXO 2030+375 (\cite{camero:2005}). 
A plausible picture may be delineated as follows. 

1) We assume that an accretion disk surrounds the NS in A0535+26. 
It is definitely the case during
the giant outbursts, as the spin-up measurements and QPO observations suggest 
(\cite{sembay:1990, finger:1996}). The disk can survive for several orbits
after the giant outburst. The observed strong spin-up during
the main part of the August/September 2005 outburst (Caballero et al. 2008
and Fig. 1) confirms the presence of the disk.

2) When the accretion rate onto the NS increases 
as it approaches orbital periastron, 
the magnetospheric radius $R_m$, determined by magnetic pressure
balancing plasma pressure, decreases according to the proportionality $R_m\propto \dot M^{-2/7}$. 
Accretion occurs when the stopping radius of matter $R_a$, generally of the order of $R_m$, becomes smaller than the corotation radius 
$R_c=(GM/\omega^2)^{1/3}$, where $M$ is the NS mass and 
$\omega=2\pi/P$ is its spin frequency.
In all models of the disk-magnetospheric interaction 
a thin boundary layer between the disk and the rotating NS magnetosphere 
does exist in which the matter becomes gradually captured by the magnetic field lines.
The width of this region is model dependent, but is of the order of the disk 
thickness (e.g., $<20c_s/\omega_K$ in \cite{anzer&boerner:1983}, or $\sim 5 c_s/\omega_K$
in \cite{lrbk:1995}; here $c_s$ is the sound speed and $\omega_K$ is the Keplerian frequency). 
So for an estimate we can assume
$\Delta l\sim h \simeq 0.1 R_a$. 
The amount of mass accumulated in this layer 
can be roughly evaluated assuming the standard $\alpha$-theory 
(\cite{shakura:1973}): 
$$
\Delta M=\rho 2\pi R_a 2h \Delta l \simeq 
(4\times 10^{19} \mathrm{g}) R_{a,9}^{7/5}\alpha^{-4/5}\dot M_{-10}^{3/5}(1-\xi)^{3/5}\,,
$$
where $R_{a,9}$ is the stopping radius in units of $10^9$~cm, $\dot M_{-10}$ is
the mass accretion rate in units of $10^{-10}$~M$_\odot$/yr, $\xi=\dot J/(\dot M\sqrt{GMR_a})$ is 
the angular mometum flux through the disk normalized to the flux from the 
stopping radius.
Here we have assumed the gas pressure to dominate over 
the radiation pressure in the boundary layer 
and the main opacity to be due to Thomson scattering. Of course, 
the disk is already not Keplerian at this radius ($d\omega/dr\to 0$ at $R_a$), but for the case 
of interest $R_a\sim R_c$  we neglect the deviation from the Keplerian motion. 
At the onset of accretion $\xi\simeq 0$, 
the viscosity parameter can be very small, $\alpha \sim 0.01$ or smaller, so the amount of matter 
in the boundary layer may be as high as $\sim 10^{21}$~g.

3) As inferred from the analysis by Baan (1979), at small accretion rates
($\sim 10^{-10} M_\odot$/yr $=10^{16}$ g/s)  
the magnetopause in A0535+26 with $P\sim 104$~s may be on the verge of a 
KS instability.
The model by Baan (1977, 1979)
has sucessfully described the statistical properties and the burst waiting time -- luminosity correlation
of type II outbursts in the rapid burster MXB 1730-335. 
In the linear regime, the KS instability grows with time exponentially in proportion to
$\exp(\Gamma t)$ with the increment 
$\Gamma^2=kg_{\rm eff}\tanh(kz)$ (\cite{arons&lea:1976b, baan:1979}), where $k$ 
is the angular wave number, $z$ is the characteristic
height above the interface and $g_{\rm eff}$ is the effective 
gravity acceleration. $g_{\rm eff}$   takes into 
account the gravitational attraction 
($\propto 1/R_m^2$),  
minus stabilizing contributions from the 
centrifugal force 
($\propto \omega^2R_m$), 
the bending of field lines by the disk, 
the curvature of the magnetic field near the magnetopause 
($\propto 1/(\rho_m R_m^7)$), 
possible velocity gradients in the plasma flow, etc. 
The condition 
for the instability to occur is the change of the sign of $g_{\rm eff}$ from negative (stability)
to positive (instability), which defines the zero-surface $g_{\rm eff}(R_z)=0$ at some
distance from the NS.
We note that under the assumption of the corotation of matter and the 
magnetosphere, the instability can occur only when the zero-surface is  
inside the corotation radius, $R_z<R_c$, because of the centrifugal term in $g_{eff}$.
This implies that instability can develop
only when accretion is permitted centrifugally. This is consistent with the observation in A0535+26 of flaring activity, and accretion.

As the magnetopause moves inwards through 
the zero-surface with increasing $\dot M$, the simplified 
analytical treatment by Baan (1979) shows that 
the lowest harmonics are the first to become unstable.
This finding however remains controversial.  
Early numerical simulations of the disk-magnetosphere interaction 
(e.g. Rast\"atter \& Schindler 1999) 
have shown that the high-$m$ modes grow faster.
The contribution of the $B_\phi$ component of the field 
in the disk can suppress high-$m$ modes. 
Simulations by Wang \& Nepveu (1983) have shown
that the high-$m$ modes grow faster but may later merge, 
forming larger filaments. 
The first global 3D MHD simulations of the disk-magnetosphere 
interaction through the KS instability (Romanova et al. 2007)
have shown that the low-$m$ modes dominate for a wide range of parameters. 
It is not however clear which factors are responsible for suppressing the high-$m$ modes.
These 3D simulations have shown 
that at small accretion rates matter accretes through funnel streams, 
while at large accretion rates, matter accretes through 
gravitational KS instabilities.
The 3D simulations have been carried out for quite small magnetospheres ($\sim 10 R_{NS}$)
which are more applicable for the millisecond pulsars. 
This is probably why the accumulation of matter at small $\dot M$ and bursting events have not yet been observed in these simulations.
Therefore, we can assume that for Be/X-ray transients with large magnetospheres, such as A0535+26, low-mode magnetospheric instability, initiated by an increase in $\dot M$, can develop.

In such binary systems the infall of plasma
blobs from the magnetopause can also be triggered by variation of the 
rate of mass captured from the variable stellar wind. 
So short X-ray flares or spikes can in principle be observed at 
any orbital phase. An "unexpected" X-ray flare in A0535+25 observed by INTEGRAL 
in October 2003 during the prolonged
quiescence period (\cite{hills:2007}) could be an example of such an event.

4) The low-mode magnetospheric instability can break-up  
the entire boundary layer in the disk. 
All material stored in the boundary layer may then rapidly enter the magnetosphere. Our estimate above
indicates that up to $\sim 10^{21}$~g can fall onto the NS surface on the 
free-fall time scale (of the order of $20$ s) producing short
X-ray spikes with a maximum X-ray luminosity of up to  $\sim 5\times 10^{36}$ erg/s
(\textit{feature 1 from the Introduction}). 
The amount of matter falling onto the NS surface during the spikes in 
the considered outburst of A0535+26 can be estimated from the light curve (Fig. 1) and is found to be 
$\Delta M\simeq \int\dot M dt\lesssim 10^{21}$~g in each spike. The characteristic 
timescale of the spikes is $\sim 10^4$ s and agrees with the time required to replenish 
the boundary layer $\sim \Delta M/\dot M$. 
The angular momentum supply from the disk during flares changes the NS spin period 
by the fractional amount $|\Delta P/P|=\Delta M\sqrt{GMR_a}P/(2\pi I)\lesssim
7\times 10^{-6}$ where we assume that the NS moment of inertia is $I=10^{45}$ g cm$^2$ and 
$R_a=R_c=10^9$ cm. This is within the uncertainty of the period measurement 
($\Delta P/P\sim 5\times 10^{-6}$, \cite{caballero:2008}), which explains
the lack of noticeable spin-up of the NS at the initial phase of the outburst
(\textit{feature 2 from the Introduction});
see also the BATSE observations of A0535+26 presented in (\cite{bildsten:1997}),  and 
of GS 1843-02 in \cite{finger:1999}.

5) The plasma 
entering the magnetosphere via the low-mode KS instability may become frozen 
into the magnetic field more close to the NS than the main accretion flux and hence fall 
along different magnetic field lines than those by which the quasi-stationary
accretion is channeled. This explains the observed difference in the cyclotron line 
energy during the initial X-ray spike and the remainder of the outburst 
(\textit{feature 3 from the Introduction}). 
Similarly to Her X-1 (Staubert et al., 2007), the absence of a radiation-dominated accretion column in A0535+26 is suggested by the independence of the CRSF energy late in the outburst from the observed X-ray luminosity (\textit{feature 4 from the Introduction}). 
The height of 
the emission region above the NS surface is then about several hundred meters. 
To change the CRSF energy by the fractional 
amount of $\Delta E_c/E_c\sim 10\%$, the emission region during the spike
is required to move towrds the NS surface by the amount 
$\Delta R/R_{NS}\sim 3\%$ (assuming  
the dipole magnetic field), i.e. by about 300 meters. Therefore, 
the emission from the spike is likely to originate 
very close to the NS surface. 
This explains the different pulse profiles during the spike, 
especially in hard X-rays (\textit{feature 5 from the Introduction}). 

The smooth change in pulse profile shape at energies above the cyclotron 
resonance (as in Her X-1, Klochkov et al. 2008) 
may correspond to a pencil-beam diagram of emission during the flare. The photon cross-section may then change smoothly, at the cyclotron resonance energy, in the strong magnetic field along the direction of the field. In the main outburst, however, the accretion column is higher (see above), the density increases such that it can 
become optically thick in the direction normal to the field,
and if there is a temperature gradient to the center, 
a fan-like beam can be additionally formed by extraordinary 
photons\footnote{Ordinary polarized photons have electric field vector {\bf E}
lying in the plane formed by the magnetic field and the wave vector of 
the photon; the {\bf E}-vector of extraordinary photons is perpendicular to this plane.}. 
The cross-sections of extraordinary photons far from the resonance 
($E\ll E_c$) are proportional to $\sigma_\perp\simeq \sigma_T(E/E_c)^2$, which is a far smaller cross-section than that of ordinary photons 
$\sigma_\parallel \simeq \sigma_T(\sin^2\theta+
\cos^2\theta(E/E_c)^2)$. The e-photons are then able to escape effectively from larger optical depth. (Here $\sigma_T$ is the Thomson scattering 
cross-section and $\theta$ is the angle between the incident photon 
and the magnetic field; see 
Harding \& Lai, 2006\footnote{A similar dependence holds for 
free-free absorption cross-sections (Kaminker et al. 1983).}). 
Above cyclotron resonance, 
$\sigma_\perp\simeq \sigma_\parallel\simeq \sigma_T$ 
and photons escape from
small optical depths. This can explain the disappearance
of the large hump at pulse phase $\sim 0.7$ after 
crossing the CRSF energy (Fig. 3 in Caballero et al. 2008).

6) As the accretion rate increases, the NS starts spinning-up and the angular momentum
flux through the inner disk rapidly approaches its maximum possible value 
$\sim \dot M\sqrt{GMR_a}$
(see Caballero et al. 2008 and Fig. 1). 
The parameter $\xi$ in the expression for $\Delta M$ above increases from
zero to almost one, which strongly reduces the mass in the disk boundary layer and hence 
the mass available for unstable accretion. 
The amplitude of spikes rapidly decreases and we do not
see strong flaring activity during the spin-up stage 
(\textit{feature 6 from the Introduction}); 
see also Figs. 2 and 10 in Finger et al. 1999
for the case of GS 1843-02. This consideration suggests that flares can reappear after 
the spin-up stopping, for which there is indeed a hint in the 
%Swift BAT 
light curve of the A0535+26 
outburst during September 13-15, 2005. At this time the NS had stopped 
spinning-up (see Fig. 1).

\section{Discussion}

The magnetospheric instability model proposed for the pre-outburst flares in A0535+26 
is generic and can be applied to other transients. The prerequisite,
however, is that the source must be on the verge of low-mode instability, 
which depends on the NS magnetic field, the spin period, 
the accretion rate and possibly other parameters 
(e.g. the misalignment of the magnetic dipole and/or NS spin axis
relative to the orbital angular momentum). It is possible that 
the transition from propeller to accretion stage can occur 
without a strong KS instability being observed. 
The model predicts 
changes in pulse profiles and in the cyclotron line
energy during the short flares, as observed in A0535+26 (\cite{caballero:2008}), 
which can be compared with observations of other sources. 

Are there other possible explanations for the observed pre-outburst X-ray spikes? 

The SPH modeling of an accretion disk surrounding the NS (Hayasaki \& Okazaki, 2006) reproduces normal outbursts at successive periastron passages. In some instances the modeling shows a single peak preceding outburst maximum. The accretion disk is formed from the Roche lobe overflow of the coplanar circumstellar disk surrounding the Be companion star. The important feature of these SPH 
simulations is the transient, one-armed spiral structure of 
the accretion disk, induced 
by a phase-dependent mass accretion.  
Such accretion disks can  
be responsible for mass transfer enhancement close to periastron. 
The model by Hayasaki \& Okazaki, however, ignores the disk-magnetosphere interaction critical to transient accretion onto magnetized NS, and can not reproduce the observed flaring activity.

Spruit \& Taam (1993) also found a viscous, disk-magnetosphere instability that was associated with the magnetospheric boundary 
around the corotation radius. This instability 
results in a cyclic enhancement of the mass accretion rate on the viscous time scale at 
the magnetospheric boundary. While this timescale can be as short as $10^4$ s for the corotation
radius $10^9$ cm, the chaotic behaviour of flares observed in A0535+26 and, most importantly, 
other features (different pulse profiles and CRSF energy, and absence of the flaring during the NS spin-up
phase) imply that this model can not provide the correct explanation.

Our model for short flares observed at the onset of outbursts 
in the transient X-ray pulsar A0535+26 is based on 
magnetospheric instability. It succesfully explains all 
features of the outburst observed during  August-September 2005 by RXTE and INTEGRAL, and makes
clear predictions for future observations. The present-day accuracy 
of cyclotron line measurements in accreting neutron stars, precise timing analysis and
evolution of X-ray pulse shapes with luminosity at different energies, have became 
the working tools to probe the non-stationary accretion onto magnetized neutron stars.  

\begin{acknowledgements}
The authors acknowledge H. Spruit, F. Meyer, R. Sunyaev and V. Suleimanov for useful discussions. 
KAP thanks the staff of the IAAT for hospitality and 
grants DAAD A/07/09400 and RNP-2.1.15940 for support.
Research has made use of \textit{Swift}/BAT transient monitor results provided by the \textit{Swift}/BAT team. 
\end{acknowledgements}

\end{document}